\documentclass[%
 reprint,
superscriptaddress,
 amsmath,amssymb,
 aps,
prb,
]{revtex4-2}

\usepackage{textgreek}
\usepackage{booktabs}
\usepackage{array}
\usepackage{multirow}
\usepackage{amsmath, amssymb, amsthm}
\usepackage{mathtools}
\usepackage{bbm}
\usepackage{graphicx}
\usepackage{physics}
\usepackage{tensor}
\usepackage{siunitx}
\usepackage[version=4]{mhchem}
\usepackage{tikz}
\usepackage{xcolor}
\usepackage{listings}
\usepackage{autobreak}
\usepackage[colorlinks,unicode]{hyperref} 
\usepackage{xurl}
\usepackage[most]{tcolorbox}

\usepackage{xr}
\usepackage{prettyref}

\lstdefinestyle{console}{
    basicstyle=\footnotesize\ttfamily,
    breaklines=true,
    postbreak=\mbox{\textcolor{red}{$\hookrightarrow$}\space}
}

\newrefformat{fig}{Fig.~\ref{#1}}
\newrefformat{tbl}{Table~\ref{#1}}
\newrefformat{appx}{Supplementary Material Sec.~\ref{#1}}

\usetikzlibrary{calc}
\tikzset{every picture/.style={line width=0.3pt}} 


\DeclareSIUnit{\au}{a.u.}

\newcommand*{\ke}{\vb{k}_{\text{e}}}

\newcommand*{\khi}[1]{\vb{k}_{\text{h#1}}}

\newcommand*{\re}{\vb{r}_{\text{e}}}

\newcommand*{\rhi}[1]{\vb{r}_{\text{h#1}}}

\newcommand*{\me}{m_{\text{e}}}
\newcommand*{\mh}{m_{\text{h}}}
\newcommand*{\Eg}{E_{\text{g}}}

\newcommand*{\EBT}{E_{\text{B,T}}}
\newcommand*{\EBX}{E_{\text{B,X}}}

\begin{document}

\title{What is the signature of a trion in photoemission?}
\author{Jinyuan Wu}
\affiliation{Department of Materials Science, Yale University, New Haven, CT 06520}
\author{Zachary H. Withers}
\affiliation{Department of Physics and Astronomy, Stony Brook University, Stony Brook, NY 11794-3400}
\author{Thomas K. Allison}
\affiliation{Department of Physics and Astronomy, Stony Brook University, Stony Brook, NY 11794-3400}
\affiliation{Department of Chemistry, Stony Brook University, Stony Brook, NY 11794-3400}
\author{Diana Y. Qiu}
\email{diana.qiu@yale.edu}
\affiliation{Department of Materials Science, Yale University, New Haven, CT 06520}

\begin{abstract}
    Recent advances in time- and angle-resolved photoemission spectroscopy (tr-ARPES) allow for the probing of multiparticle excited-states in reciprocal space. While neutral two-particle excitations (excitons) have been observed in tr-ARPES, signatures of 
    trions---three-quasiparticle bound states---have only been probed via optical spectroscopy.
    Here, we develop a general theory for the ARPES signature of trions in the model system of a monolayer transition metal dichalcogenide (TMD).
    We simulate the ARPES signals of both positively and negatively charged trions and show that the interaction of the residual holes, or electron and hole, lead to large energy shifts, on the order of the exciton binding energy, compared to the exciton  signal. 
    For positive trions, the additional momentum degree of freedom of the residual particles removes any strict lower bound on the photoemission energy, leading to distinctive asymmetric spectral features.
    For negative trions, the photoemission process causes the tr-ARPES spectrum to reproduce inverted images of the exciton band structure for multiple exciton states, encompassing both spin-allowed and spin-forbidden states, providing a direct momentum-resolved probe of both trion and exciton physics.
\end{abstract}

\maketitle


Time- and angle-resolved photoemission spectroscopy (tr-ARPES) probes ultrafast non-equilibrium electron dynamics in condensed matter systems by first exciting the system with an ultrashort pump pulse and then recording photoemission initiated by a probe pulse as a function of delay time between the pump and the probe \cite{boschini2024time}.
Various non-equilibrium phenomena have been explored by tr-ARPES, including non-equilibrium occupation of surface states in a topological insulator \cite{neupane2015gigantic,ciocys2020manipulating} and Floquet bands \cite{mahmood2016selective}.
Recently, advancements in tr-ARPES have enabled the direct measurement of the electron constituents of excitons \cite{madeo2020directly,ma2022multiple,mori2023spin,kunin2023momentum,guo2025moir}.
In principle, tr-ARPES should also be able to probe multi-particle excitations other than excitons.

In absorption and photoluminescence (PL) measurements of doped transition metal dichalcogenides (TMDs), doping results in
a splitting of the first exciton peak, which can be interpreted as the formation of three-particle bound states, known as trions, in the dilute doping limit
\cite{mak2013tightly,koperski2017optical,yoon2021,perea2022exciton}.
Additionally, in conventional ARPES, a renormalization of the valence band has been attributed to trion formation after photoemission creates a photo-hole \cite{katoch2018giant}.
These measurements do not, however, directly probe the electron (or hole) constituents of the trion, unlike what has been achieved for excitons in tr-ARPES \cite{madeo2020directly,ma2022multiple,mori2023spin,kunin2023momentum,guo2025moir}. 
Resolving the ARPES signal of a trion could provide unambiguous momentum-resolved signatures of three-body bound states, going beyond what purely optical methods can achieve, and could also shed light on the existence of other multi-particle excitations at higher doping, including Fermi polarons and Suris tetrons \cite{koudinov2014suris,efimkin2017many,sidler2017fermi}.

In this paper, we present a theoretical framework to understand trion signatures in tr-ARPES.
We use a two-band model built on top of Bethe-Salpeter equation (BSE) calculations of trions and excitons \cite{druppel2017diversity,arora2020dark,qiu2015nonanalyticity,deslippe2012berkeleygw} in monolayer \ce{MoS2}, a prototypical 2D semiconductor. We find clear qualitative differences between the signatures of positive and negative trions and excitons. 
While excitons appear as either a replica of the valence band or a replica of the conduction band, depending on the temperature, as established in Refs. \cite{rustagi2018photoemission,madeo2020directly,chan2023giant}, trions give rise to several distinctive spectral features arising from the interaction of the residual quasiparticles. 
First, unlike optical spectroscopy, where the shift of peak positions due to trion formation is small, the interaction of residual particles results in a shift of the spectral features comparable to the magnitude of the exciton binding energy (hundreds of meV in monolayer TMDs~\cite{qiu2013optical,Qiu2016screening,Ugeda2014giant,Ye2014}), well within the resolution of modern electron spectroscopy experiments. 
Analogous to the ionization of the atomic hydrogen anion H$^-$, the energy required to remove an electron from the negative trion is small, so that the negative trion photoemission features appear just a few 10's of meV below the conduction band. 
For the positive trion, analogous to ionizing the hydrogen molecular cation H$_2^+$, the electron is more strongly bound, and photoelectrons appear at lower energy than the exciton case.
Second, for the positive trion, the combination of Coulomb repulsion of the two unbound holes and the momentum content of the trion wave function lead to a distinctive asymmetric broadening of the ARPES signals,
while for the negative trion, the center-of-mass (COM) dispersion of the entire Rydberg series of the residual exciton is observed in the ARPES signals, providing a unique method to observe the dispersion of excitons.
Our work establishes a direct route to detecting and characterizing trions experimentally in momentum space, while also showing that photoemission from trions can serve as a unique probe of \textit{exciton} bandstructure, whose selection rules directly reflect the internal composition of the trion wavefunction.


The ARPES signal is given by a sum of contributions of signals from different excited states, each of which is given by a Fermi golden rule between an initial state $\ket{\Psi_{n}} \coloneqq |\Psi_{S'\vb{P}}\rangle$ and a final residue state $\ket{\Psi^{\text{residue}}_{m}} \coloneqq |\Psi_{S \vb{Q}}^{\text{residue}}\rangle$ 
\begin{equation}
    \begin{aligned}
        I_{S' \vb{P}}(\vb{k}, \omega, t) &\propto  \sum_{c m} \abs{M_{f c \vb{k}}}^2 \abs{\mel{\Psi_{m}^{\text{residue}}}{ c_{c \vb{k}}}{\Psi_{S' \vb{P}}}}^2  \\
        &\quad\quad \times \delta(\omega - E_{n } + E_{m }^{\text{residue}}),
    \end{aligned}
    \label{eq:tr-arpes-incoherent-final}
\end{equation}
where $\vb{k}$ is the momentum of the photoemitted electron such that $\vb{P} = \vb{k} + \vb{Q}$; 
$S, \vb{Q}$ are the primary quantum numbers of a generic final state in a periodic system;
$\abs{\mel{\Psi_{m }^{\text{residue}}}{ c_{c \vb{k}}}{\Psi_{n }}}^2$ is the spectral weight,
whose intensity depends on the overlap of the initial and final state wave functions;
and $M_{fc\vb{k}}$ is the dipole transition matrix element from the conduction band to a high-energy outgoing state.
In this work, we focus on the spectral function of the trion, which gives the pole structure of $I(\vb{k}, \omega, t)$, and the influence of the matrix element $\abs{M_{f c \vb{k}}}^2$ on the intensity is ignored.
$E^{\text{residue}}$ denotes the energy of the interacting residue of the trion state after photoemission removes one electron.

\begin{figure}
    \centering
    \includegraphics[width=\linewidth]{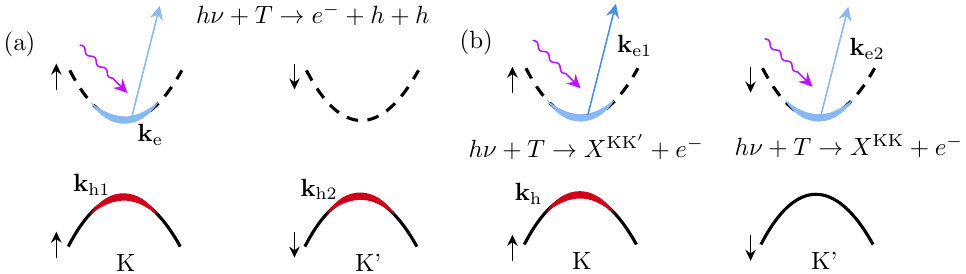}
    \caption{Schematic of two trion modes and photoemission processes considered in this paper.
    (a) The positive trion, in which there is only one electron which can be photoemitted and detected by ARPES.
    (b) The negative trion mode, with two electrons, either of which can be photoemitted in ARPES, leaving behind either an intervalley K-K' or intravalley K-K exciton.}
    \label{fig:schematic}
\end{figure}

Throughout this work, we primarily focus on the lowest-energy optically active positive or negative trion, which in monolayer \ce{MoS2} has the structure of an intravalley spin-allowed A-series exciton in one valley bound to an additional quasiparticle in the opposite valley, as illustrated in \prettyref{fig:schematic}~\cite{berkelbach2013theory,druppel2017diversity}.
Because of the band structure and spin-valley locking in monolayer \ce{MoS2}, the electron-hole pair in the same valley share the same spin, which is opposite to the spin of the additional quasiparticle in the opposite valley (\prettyref{fig:schematic}).
Below, we consider photoexcitation in the K valley. 

We describe the trion within a simple two-band model, which consists of one valence and conduction band of the same spin in the K/K' valleys. 
A two-band model describing the positive trion mode in monolayer \ce{MoS2} reads 
\begin{equation}
    \begin{aligned}
        H &= \frac{\ke^2}{2\me} + \Eg 
        + \frac{\khi{1}^2}{2\mh} + \frac{(\khi{2} - \vb{w})^2}{2\mh} \\
        &\ \ + V(\rhi{1} - \rhi{2}) - V(\re - \rhi{1}) - V(\re - \rhi{2}).
    \end{aligned}
    \label{eq:ehh-two-band}
\end{equation}
Here, we are using atomic Hartree units, and for simplicity, we shift the center of the coordinate system to the K valley, with a bandgap $\Eg$.
$\vb{w}$ represents the vector from K to K'.
$\vb{k}_{\text{e,h1,h2}}$ represents the deviation of a quasiparticle (the electron (e) or either hole (h1,h2)) from K in momentum space.
It can be easily seen that the total momentum of the trion $\vb{P} = \khi{1} + \khi{2} + \ke$ is conserved, and we can rewrite the Hamiltonian in terms of 
the total momentum, which should only appear in the total kinetic energy term,
and two internal degrees of freedom.
Under an appropriate coordinate transform similar to that in Ref.~\cite{rustagi2018photoemission}, \eqref{eq:ehh-two-band} can be recast into 
\begin{equation}
    \begin{aligned}
        H &=  \frac{(\vb{P} - \vb{w})^2}{2 M} + \Eg \\ 
        &\ \  + \underbrace{
            \frac{\vb{k}_1^2 + \vb{k}_2^2}{2\mu} + \frac{\vb{k}_1 \cdot \vb{k}_2}{\me}
            + V(\vb{r}_1 - \vb{r}_2) - V(\vb{r}_1) - V(\vb{r}_2)
        }_{H_{\text{internal}}},
    \end{aligned}
    \label{eq:internal-ham}
\end{equation}
where $\mu = \me \mh / (\me + \mh)$, and $\vb{k}_{1,2}$ are COM momenta of the holes.
The lowest-energy configuration corresponds to a trion COM momentum $\vb{P}=\vb{w}$.
We note that the trion has one more internal momentum variable than the exciton does,
which has consequences in the spectroscopic behavior of trions.
A similar procedure can be performed for the negative trion.
The two two-band models can then be solved by a variational ansatz inspired by the Helium atom.
The signatures of positive and negative trions in TR-ARPES are then determined from Eq. \eqref{eq:tr-arpes-incoherent-final}.




\begin{figure}
    \centering
    \includegraphics[width=\linewidth]{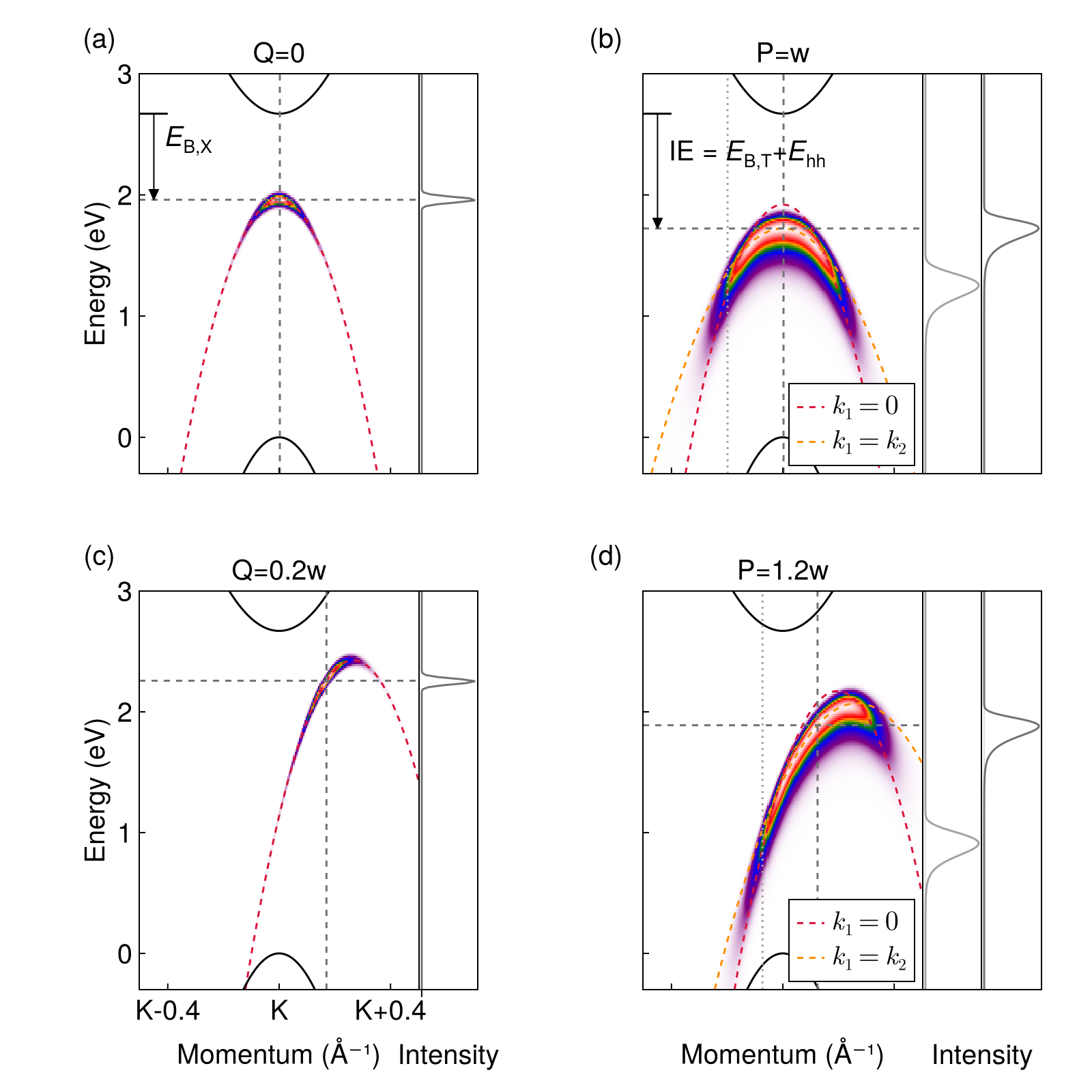}
    \caption{Comparison of signals from the lowest energy optically active exciton and positive trion in the one-electron emission spectrum in tr-ARPES of monolayer \ce{MoS2}. 
    (a) ARPES heatmap of the exciton at $\vb{Q} = 0$ and (b) the trion at $\vb{P} = \vb{w}$. 
    The momentum path is along K-K'. The x-axis corresponds to $\vb{k}$, the momentum of the photoemitted electron.
    The solid black curves are the valence and conduction bands.
    (a) The exciton plot.
    The red dashed curve highlights the dispersion of the maximum ARPES intensity.
    The panel on the right is an energy distribution curve (EDC) showing the intensity along the gray dashed vertical line in the heatmap.
    (b) The trion plot. The red dashed curve is a replica of the valence band,
    which approximately follows the dispersion of the maximum ARPES intensity.
    The orange dashed curve follows the upper edge of the signal
    and has an effective mass of $2\mh$.
    The two panels on the right correspond to the EDC  
    at K (dark gray dashed line) and away from K (light gray dotted line).
    (c,d) Similar to (a,b), but for the $\vb{Q} = 0.2 \vb{w}$ exciton and the $\vb{P} = 1.2 \vb{w}$ trion, respectively.
    The horizontal dotted line in all the four heatmaps give the energy of the global maxima of the ARPES intensity.}
    \label{fig:exciton-ehh-comparison}
\end{figure}


We start by analyzing the positive trion composed of two holes and one electron.
In monolayer \ce{MoS2}, the lowest energy optically active positive trion has one hole in each valley so that the total momentum is $\vb{P} = \vb{w}$.
The photoemission process is illustrated in \prettyref{fig:schematic}(a). In \prettyref{fig:exciton-ehh-comparison}, we compare the TR-ARPES signatures of the zero-momentum exciton in the K valley and the $\vb{P} = \vb{w}$ positive trion.

The exciton signal (\prettyref{fig:exciton-ehh-comparison}(a,c)) is red-shifted by the exciton binding energy $\EBX$ with respect to the conduction band and
appears as a replica of the valence band \cite{rustagi2018photoemission}. The position of the exciton feature can be understood from the photoemission process $h\nu + X \rightarrow e^{-} + h$: the initial exciton ($X$) is photoionized, and hence, the photoemission signal appears at the ionization energy (or binding energy) of the exciton measured with respect to the unbound electron at the conduction band edge. For the positive trion($T$), the photoemission process is $h\nu + T \rightarrow e^{-} + (h+h)$, corresponding to the ionization of a trion leaving behind a residue of two interacting holes. The ionization energy (IE) is equivalent to the sum of the binding energy of the trion ($\EBT$) -- defined as the energy difference between three noninteracting quasiparticles and the energy of the trion formed by them -- and the energies of the interacting residual holes ($E_{\text{hh}}$). 
The energy of the residual holes in the COM frame, $E_{\text{hh}}$ is not a single number but rather a distribution of energies arising from the continuum of Coulomb repulsion between the holes. A precise calculation of this interaction requires rigorous consideration of the non-trivial dynamics of two holes during the photoemission process. Here, we account for the effect of this interaction by introducing an energy-dependent weight in Eq. \eqref{eq:tr-arpes-incoherent-final} that reflects a classical estimate of the relative phase space or the likelihood for two repulsive charges to form a  given two-hole configuration.
This weight factor with respect to $E_{\text{hh}}$ leads to a significant redshift of $\sim$\SI{0.2}{eV} of the trion signal compared with that of the exciton,
in addition to the redshift due to $\EBT - \EBX \simeq \SI{0.04}{eV} \ll \EBX$ \cite{druppel2017diversity}.


Energy conservation (with respect to the full distribution of $E_\text{hh}$) imposes a lower bound on the energy of the residual two-hole state, which in turn defines \textit{maximum} photoelectron energies. In contrast, there is no analogous lower bound on the photoelectron energy.
These different constraints on the upper and lower photoelectron energy combine to produce a broad, inherently asymmetric line shape in the trion  ARPES signal (\prettyref{fig:exciton-ehh-comparison}(b)), independent of lifetime and probe profile.
Thus, the onset of trion formation in TR-ARPES is characterized by both a large redshift and an asymmetric lineshape compared to the exciton.
A similar asymmetric line shape is also observed for both negative and positive trions in PL due to the continuum of the energy of the remaining electron or hole after optical recombination \cite{christopher2017long}.
However, the physical process behind PL is different ($T \to h + h \nu$), corresponding to the recombination of an electron and hole leaving behind a single residual hole. Consequently, the relevant  term that comes into the energy conservation is limited to $\EBT$, leading to a small splitting of the trion and exciton peaks on the order of tens of meV~\cite{mak2013tightly,koperski2017optical,yoon2021,perea2022exciton}. 

The continuum of residue state energies implies that it is no longer possible to describe the trion ARPES feature with a single effective-mass dispersion.
The asymmetric ARPES feature exhibits two distinct curvatures.
The upper bound (broadened by the continuum of $E_\text{hh}$) follows a dispersion with an effective mass of $2\mh$ 
shown by the orange dashed curves in \prettyref{fig:exciton-ehh-comparison}.
The curvature associated with the maximum signal intensity is, however, different.
The maximum intensity at a given photoelectron momentum is controlled by the overlap between the trion wavefunction spectral weight and the manifold of $(\khi{1},\khi{2})$ satisfying energy conservation. When the trion spectral weight is sufficiently localized in momentum space, the dominant contribution comes from configurations where one of the hole's velocities is the same as the velocity of the trion, i.e., where either $\vb{k}_1 \approx 0$ or $\vb{k}_2 \approx 0$. Imposing this condition while enforcing energy conservation yields a second dispersion relation with an effective mass that is roughly $\mh$. This dispersion, highlighted by the red dashed curve in \prettyref{fig:exciton-ehh-comparison}, approximately follows the locus of maximum photoemission intensity. Thus, the trion ARPES feature naturally exhibits a leading-edge effective mass $2\mh$ and an effective mass $\sim \mh$ associated with the intensity peak.
The peak of the signature appears at 
\begin{equation}
    \vb{k} = \frac{\me}{M} (\vb{P} - \vb{w}), \  
    \omega = \frac{\me}{2M^2} (\vb{P} - \vb{w})^2 + \Eg - \EBT - E_{\text{hh}},
    \label{eq:ehh-center}
\end{equation}
where $M = 2\mh + \me$.

The above discussion on effective masses and intensity peak positions is general and remains valid when the trion COM momentum, $\vb{P}$, deviates from $\vb{w}$ (\prettyref{fig:exciton-ehh-comparison}(d)).
We compare the ARPES signatures of a finite-momentum exciton and trion. For an exciton with COM momentum $\vb{Q}$, the center of the ARPES replica follows \cite{rustagi2018photoemission}
\begin{equation}
    \vb{k} = \frac{\me}{M} \vb{Q}, \quad \omega = \frac{\me}{2M^2} \vb{Q}^2 + \Eg - \EBX.
    \label{eq:ex-center}
\end{equation}
where $M$ is now the total exciton mass. Comparing Eqs.~\eqref{eq:ehh-center} and \eqref{eq:ex-center}, we see that the exciton and trion peak positions have the same functional dependence on the COM shift (either $\vb{Q}$ or $\vb{P}-\vb{w}$), but with different total masses and binding-energy offsets ($\EBX$ vs $\EBT + E_{\text{hh}}$). Therefore, when an exciton and trion have the same COM momentum shift $\vb{Q}$, their loci of maximum ARPES intensity are separated in both frequency and momentum, as illustrated in \prettyref{fig:exciton-ehh-comparison}(c,d).


Next, we turn to the ARPES signature of the negative trion. The residue after photoemission now consists of an electron and a hole that can interact. 
Materials with stable trions have strongly bound excitons,
and thus the final states after photoemitting an electron from a negative trion correspond to bound excitons formed by the residual electron and hole ($h\nu+T\rightarrow X + e^{-}$). 
A series of resonances appear in the ARPES spectrum corresponding to leaving the bound exciton in different states.
The highest energy photoemission signal corresponds to the lowest-energy 1s exction, appearing at $\text{EA}=\EBT-\EBX$ below the conduction band edge, where EA is the electron affinity of the exciton. 
This is a small shift (approximately $\SI{40}{meV}$ \cite{druppel2017diversity}), but within the resolution state-of-the-art XUV tr-ARPES setups \cite{Allison_APLPhotonics2025,Mills_RSI2019,Sie_NatComm2019}. 
The bound exciton resonances appearing in the photoemission spectrum have a rich dispersion structure corresponding to the exciton band structure. 

\begin{figure}
    \centering
    \includegraphics[width=\linewidth]{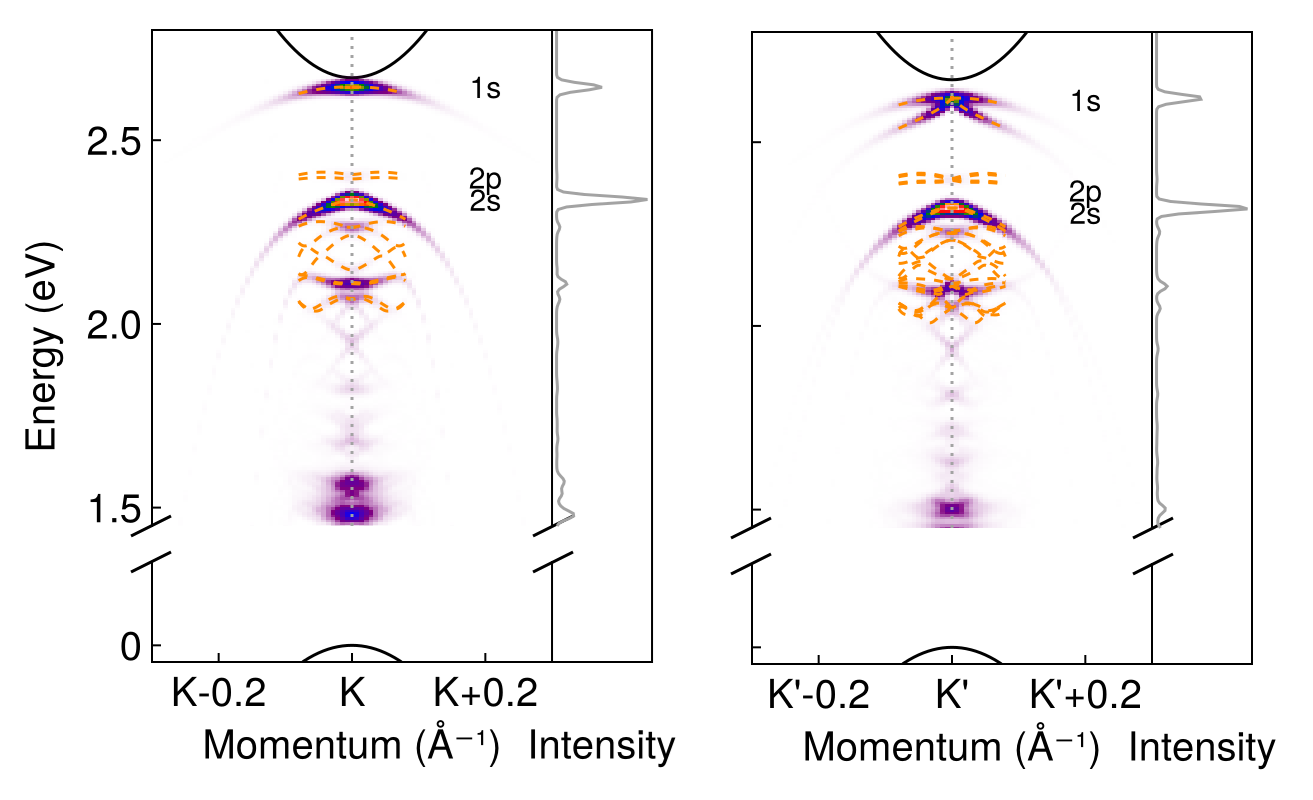}    
    \caption{One-electron photoemission spectrum from the lowest negative trion in tr-ARPES of monolayer \ce{MoS2}. The signal at the K valley is from the dispersion of the A-series of excitons with $\vb{Q} \approx \mathrm{K}$ (i.e. the intervalley exciton), while the signal in the K' valley is from the A-series of excitons with $\vb{Q} \approx 0$ (i.e. the intravalley exciton). The exciton dispersion is taken from Ref.~\cite{qiu2015nonanalyticity}.
    The orange dashed lines trace the exciton bands, the highest ones corresponding to 1s, 2p, and 2s exciton states.
    }
    \label{fig:negative-trion-exciton-final-state}
\end{figure}

We calculate the photoemission intensity of the negative trion in monolayer \ce{MoS2} using Eq.~\eqref{eq:tr-arpes-incoherent-final}, taking the residual excitons from a previous \textit{ab initio} GW-BSE calculation in Ref.~\cite{qiu2015nonanalyticity}. The resulting tr-ARPES spectrum
for the $\vb{P}=\vb{w}$ trion is shown in \prettyref{fig:negative-trion-exciton-final-state}. Unlike the positive trion, the negative trion has photoemission signals from both valleys because the trion has one electron in each valley, either of which can be photoemitted. Not all exciton states are visible in photoemission.
The spectral weight is only non-zero for exciton final states that have the same electron and hole character as the original trion. 
The structure of the $\vb{P}=\vb{w}$ negative trion is shown in \prettyref{fig:schematic}(b) and consists of an exciton in K and an additional charge of opposite spin in the lowest conduction band in K'.
If an electron is photoemitted from the K valley, a spin-forbidden intervalley exciton is left behind (\prettyref{fig:schematic}(b)), and the ARPES intensity is simply the inverted dispersion of the $\vb{Q}\approx \mathrm{K}$ exciton bands, of which only the intervalley spin-forbidden A-series excitons have significant spectral weight. (Here, we use spin-forbidden(allowed) to denote the  excitons composed of transitions between valence and conduction bands of the opposite(same) spin.) The spin-orbit split B series and $\Gamma$-K excitons do not contribute.
There are also selection rules within the Rydberg series. Within the variational ansatz, the lowest trion state has the same s-like angular momentum as the exciton 1s states \cite{berkelbach2013theory}. Within the current photoemission formalism, the spectral weight imposes a selection rule: it is only non-zero for residual exciton states with s-like character, and p-like excitons do not contribute to the ARPES spectrum 
Interestingly, the 2s exciton contributes more than the 1s exciton to the trion spectral function, reflecting the more delocalized nature of the trion.

If the electron is photoemitted from the K' valley, the residue is a spin-allowed intravalley exciton (20 meV lower in energy than the K valley signal due to the exchange splitting of excitons of different spin), and the ARPES spectrum is simply the inverted dispersion of the spin-allowed A-series intravalley $\vb{Q} \approx \mathrm{\Gamma}$ excitons.
For the $1s$-states in the Rydberg series, the lowest-energy exciton splits into a parabolically dispersing transverse branch and a linearly dispersing longitudinal branch \cite{yu2014dirac,qiu2015nonanalyticity,wu2015exciton,cudazzo2016exciton}. While the transverse branch can be measured optically, the longitudinal branch has only been experimentally measured very recently in momentum-resolved electron energy loss spectroscopy (Q-EELS) in monolayer hexagonal boron nitride~\cite{liu2025direct}. Remarkably, since both branches contain electron and hole states of the same character, both longitudinal and transverse exciton dispersion is visible in \prettyref{fig:negative-trion-exciton-final-state}. Thus, our calculations suggest that tr-ARPES may be an alternate way to access exciton dispersion of the entire exciton Rydberg series with finer momentum resolution than easily accessible in Q-EELS or optics.

In summary, we derive a general theory of the tr-ARPES signature of composite quasiparticles and use it to explore the tr-ARPES signatures of trions in a two-band model of monolayer \ce{MoS2} parametrized by \textit{ab initio} trion and exciton BSE calculations. Our calculations reveal that positive and negative trions give rise to characteristic spectral features that are qualitatively distinct from each other and from excitons. 
Photoemission from a positive trion leaves behind an unstable two-hole complex, resulting in a dramatic redshift and an asymmetric lineshape with a tail encompassing the energy continuum of the residual holes. Photoemission from a negative trion, on the other hand, leaves behind a stable exciton, forming a series of resonances in the photoemission spectrum, appearing as an inverted exciton band structure. 
In this way, ARPES of negative trions projects the trion wave function onto the basis of exciton states, and this projection can be used to determine the trion wave function.
Our model can be easily generalized to other multi-quasiparticle excited states and their dynamics.

In the process of revising this paper, we have become aware of another work focused on the signatures of negative trions in ARPES\cite{meneghini2025arpessignaturestrionsvan}.

\textit{Acknowledgments} --- This work was primarily supported by the National Science Foundation (NSF) Condensed Matter and Materials Theory (CMMT) program under Career Grant Number DMR-2337987. Z.W. and T.K.A. were supported by the Department of Energy, Office of Science, Basic Energy Sciences under Award No. DE-SC0022004 and the AFOSR under Award No. FA9550-20-10259. Z.H.W. acknowledges support from the U.S. National Science Foundation Graduate Research Fellowship Program. Development of the BerkeleyGW code was supported by Center for Computational Study of Excited-State Phenomena in Energy Materials (C2SEPEM) at the Lawrence Berkeley National Laboratory, funded by the U.S. Department of Energy, Office of Science, Basic Energy Sciences, Materials Sciences and Engineering Division, under Contract No. DE-AC02-05CH11231. The calculations used resources of the National Energy Research Scientific Computing (NERSC), a DOE Office of Science User Facility operated under contract no. DE-AC02-05CH11231, under award BES-ERCAP-0031507 and BES-ERCAP-0027380; and the Texas Advanced Computing Center (TACC) at The University of Texas at Austin. D.Y.Q. would like to thank Y. He for insightful discussions.

\bibliography{arpes,trion,materials}

\end{document}